\documentclass[12pt,preprint]{aastex}
\usepackage{graphicx}

\newcommand{\radec}[8][]{\ensuremath{#2^{h} #3^{m} #4 \, #5\!\!#6^{\circ} #7' #8} #1}
\newcommand{\by}[2]{\ensuremath{#1\times{}#2}}
\newcommand{\telescopesize}[2]{\ensuremath{#1}-#2}
\newcommand{\Rvalue}[0]{$207\pm35$~pc}
\newcommand{\radiusvalue}[0]{$0.12\pm0.03$ arcsec}
\newcommand{\colorindex}[2]{\ensuremath{#1\!-\!#2}}
\newcommand{\SN}[1]{SN~#1}
\newcommand{\NGC}[1]{NGC~#1}

\begin{document}

\title{A Light Echo from Type Ia Supernova \SN{1995E}?}

\author{Jason L. Quinn and Peter M. Garnavich}
\affil{Physics Department, University of Notre Dame, Notre Dame, IN 45556}

\author{Weidong Li}
\affil{Department of Astronomy, University of California, Berkeley, CA 94720-3411}

\author{Nino Panagia and Adam Riess}
\affil{Space Telescope Science Institute, 3700 San Martin Drive, Baltimore, MD 21218}

\author{Brian P. Schmidt}
\affil{Mount Stromlo and Siding Spring Observatory, Private Bag, Weston Creek P.O., Canberra ACT 2601, Australia.}

\and
\author{Massimo Della Valle}
\affil{Osservatorio Astrofisico di Arcetri, Largo Enrico Fermi 5, 50125-Firenze, Italy}

\email{jquinn@nd.edu, pgarnavi@miranda.phys.nd.edu, weidong@astro.berkeley.edu, ariess@stsci.edu, panagia@stsci.edu, brian@mso.anu.edu.au, massimo@arcetri.astro.it}

\begin{abstract}
We identify a light echo candidate from Hubble Space Telescope (HST) imaging of \NGC{2441}, the host galaxy of the Type Ia supernova 1995E. From the echo's angular size and the estimated distance to the host galaxy, we find a distance of \Rvalue{} between the dust and the site of the supernova. If confirmed, this echo brings the total number of observed non-historical Type Ia light echoes to three --- the others being \SN{1991T} and \SN{1998bu} --- suggesting they are not uncommon. We compare the properties of the known Type~Ia supernova echoes and test models of light echoes developed by Patat et al. (2005). HST photometry of the \SN{1991T} echo shows a fading which is consistent with scattering by dust distributed in a sphere or shell around the supernova. Light echoes have the potential to answer questions about the progenitors of Type~Ia supernovae and more effort should be made for their detection given the importance of Type~Ia supernovae to measurements of dark energy.
\end{abstract}

\keywords{supernovae, light echoes, individual(\object{\SN{1995E}},\objectname{\NGC{2441}})}

\section{Introduction}

Light echoes are produced by radiation scattered by dust near or along the line of sight to an astrophysical explosion. Initially seen in Nova Per 1901 \citep{Ritchey1901} and then much later in Type~II \SN{1987A} \citep{Crotts1988}, a flourish of light echoes have recently been observed. Spectacular examples have been detected for the peculiar star V838~Mon \citep{Bond2003} and the so-called ancient or historical supernovae in the LMC \citep{Rest2005}. Even distant gamma-ray bursts have been found to produce x-ray echoes by scattering off Galactic dust \citep{Vaughan2004}. Other events include the Type~II supernovae \SN{1993J} \citep{Sugerman2002}, \SN{1999ev} \citep{Maund2005}, and \SN{2003gd} \citep{Sugerman2005,VanDyk2006} and two recent (that is, non-historical) Type~Ia supernovae: \SN{1991T} \citep{Schmidt1994} and \SN{1998bu} \citep{Cappellaro2000,Garnavich2001}. This paper reports a candidate for a third Type Ia SN light echo.

Type~Ia supernovae (SNe), in particular, are important as distance indicators for cosmology but their reliability could be improved by a better understanding of the explosion mechanism and progenitors; so it is intriguing to consider using light echoes to probe their environments. Important clues to the distribution and properties of the dust in their vicinity may be found by studying the brightness and geometry of SN~Ia light echoes. The distance between the dusty region and light source is readily calculable with a few simplifying assumptions about the dust distribution \citep{Couderc1939,Sugerman2003}. Further information such as polarization measurements could even provide independent distance measurements between the observer and the explosive event \citep{Sparks1999}. A statistical estimate of the distribution of SN~Ia relative to the structures in spiral galaxies will become possible as the sample size grows.

The peak luminosity of SN~Ia are known to correlate with host galaxy morphology \citep{Hamuy1996} and star-formation history \citep{Gallagher2005} so that the intrinsically faintest events occur in E/S0 galaxies and hosts with low current star-formation rate. The progenitors of the brighter SN~Ia in spirals appear to be associated with dusty regions and come from a young population \citep{Mannucci2005}; therefore they have a good chance of producing a light echo. A pair of Type Ia supernovae, \SN{2002ic} and \SN{2005gj}, have, in any case, recently been observed in dense circumstellar environments \citep{Hamuy2003,Prieto2005,Aldering2006}. If the mass-donating star in SN~Ia binary progenitors are giants then circumstellar dust may also be detectable through light echoes.

Here we present evidence of a third light echo generated from a recent SN~Ia event. \SN{1995E} was discovered near maximum light on Feb.~20.85 1995 UT by A. Gabrielcic \citep{Molaro1995a}. The spectral identification as a Type Ia was made by \citet{Molaro1995b}. A spectrum was obtained by \citet{Riess1997} and \citet{Riess1999} determine it was discovered near maximum light, had a typical decline-rate parameter ($\Delta{}m_{15}(B)=1.06\pm0.05$), and that it showed more than $1$ mag of visual extinction using a multi-color light-curve shape fit. \citet{Phillips1999} find a host galaxy extinction of $E(\colorindex{B}{V})=0.74\pm0.03$.

\section{The Observations}
HST SNAP program 9148 took 33 STIS images of the locations of old Type Ia supernovae in the search for new Type Ia light echoes. The STIS field of view is \by{52''}{52''} and the corrected mean plate scale of the CCD is $0.050725\pm0\farcs{}00008$ per pixel. The images were taken in direct-imaging mode (50CCD), where the bandpass window is determined by the detector response alone. The STIS 50CCD has a peak throughput percentage of $14.74$\% at $550$~nm and falls to half the peak at about $380$~nm and $820$~nm. The data products included the images and calibration data. The image pre-processing and the removal of the majority of the cosmic-ray strikes were done by the HST pipeline reductions.

The STIS data set included CR-split images of \SN{1991T}, \SN{1995E}, and \SN{1998bu}. The image of \SN{1995E} in \NGC{2441} (an SBc/Sc spiral according to SIMBAD\footnote{This research has made use of the SIMBAD database,operated at CDS, Strasbourg, France.}) was obtained on 2001-Sep-04 starting at 07:37:46 (UT) approximately $2387$ days after $V$-maximum, $\Delta{}T$. The images of \SN{1998bu} and \SN{1991T} were made on 2002-Jan-18 at 23:47:47 (UT) and 2001-Nov-23 at 04:11:14.7 (UT), $1486$ and $3857$ days past maximum, respectively. Each of these images had exposure times of $900$~sec.

Two data sets (proposal ID/PIs: 8242/Savage and 9114/Kirshner) of \SN{1998bu} using HST's WFPC2 were also downloaded from the MAST archive. The target was in the PC field. The earlier visit, 2000-Jun-19, had four $320$~sec CR-split images (i.e., eight images of $160$ sec each) in F555W and the later, 2002-Jan-10, had four $1000$~sec CR-split images in F439W and four $320$~sec CR-split images in F555W\@.

\begin{figure}
\epsscale{0.77}
\plotone{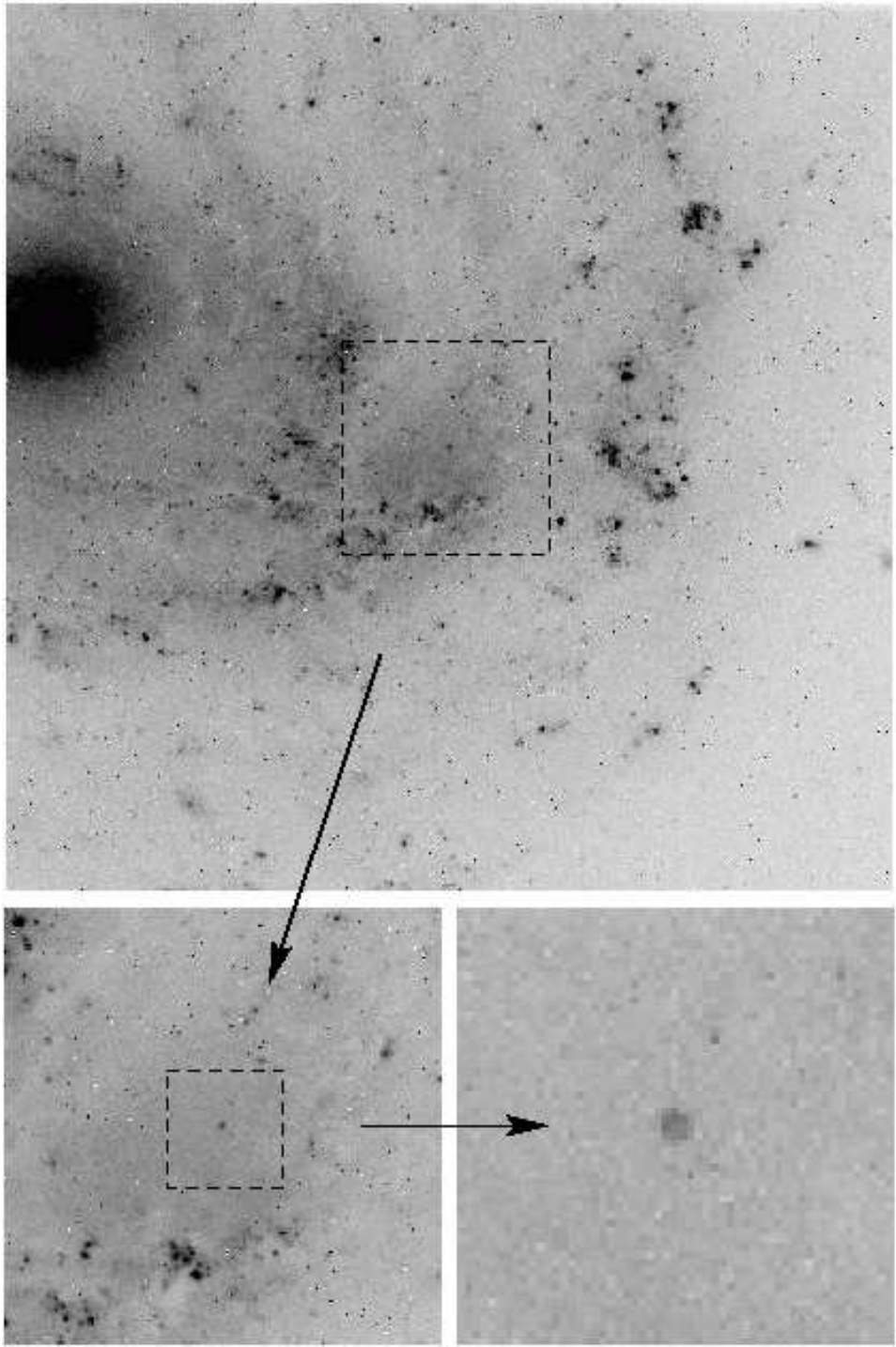}
\caption{HST STIS close-up of \SN{1995E} in \NGC{2441}. North is to the left and east is down. The source position is \radec[J2000]{07}{51}{56\fs{}5}{+}{73}{00}{35\farcs{}1} and $22\farcs{}9$ from the galactic center. The image sizes are roughly $48''\times{}48''$, $12''\times{}12''$, and $3''\times{}3''$.}
\label{fig:closeup}
\end{figure}

\section{Analysis}
A resolved source is detected in the STIS images near the position of \SN{1995E} in a region
with an otherwise smooth background suggesting that it may be a light echo. The source is in a spiral arm in the outer regions of the galaxy about $22\farcs{}9$ from the galactic center. The position of the light echo candidate (see Fig.~\ref{fig:closeup}) was compared to the position of \SN{1995E} in two ways. Both methods show the resolved source coincident with the supernova's position.

First, the pixel position of $22$ field stars and the supernova were measured from eight images obtained over four nights with the \telescopesize{1.2}{m} telescope at the Fred~L. Whipple Observatory
(FLWO) in March 1995 when the supernova was bright. The right ascension (RA) and declination (DEC) for the field stars were measured from the Digitized Sky Survey at the Space Telescope Science Institute
archive\footnote{The Digitized Sky Surveys were produced at the Space Telescope Science Institute under U.S. Government grant NAG W-2166. The images of these surveys are based on photographic data obtained using the Oschin Schmidt Telescope on Palomar Mountain and the UK Schmidt Telescope. The plates were processed into the present compressed digital form with the permission of these institutions.} and a solution computed to map the pixel positions to the celestial coordinates. The solution resulted in an RMS scatter of $0.13$ arcsec for the field stars and a position for \SN{1995E} of \radec[J2000]{07}{51}{56\fs{}54}{+}{73}{00}{35\farcs{}10}.
We also measured the centroid of the core of \NGC{2441} to have end figures of $54.72$ and $56.44$ in RA and DEC respectively. Using the world coordinate system in the HST/STIS image we measure the light echo candidate to be at
\radec[J2000]{07}{51}{56\fs{}5}{+}{73}{00}{35\farcs{}06}. This is a difference of $0.27$ arcsec in RA and $0.04$ arcsec in DEC between the suspected light echo and the position of the supernova from the ground. The difference between
the ground and STIS positions of the galaxy core is $0.44$ arcsec in RA and $0.05$ arcsec in DEC\@.
                                                                                                        
The FLWO images were obtained in good seeing and many features seen in the STIS image are visible in the ground-based image. The STIS images were convolved with a broad Gaussian and the position of seven common features were measured in pixel coordinates. Using the ``geomap'' and ``geotrans'' tasks in IRAF, the FLWO image was transformed to the STIS pixel coordinates with an RMS scatter of $0.18$ arcsec. The difference in position between the supernova in the transformed FLWO image and the echo candidate in the STIS image is $0.16$ arcsec, which is within the error of the measurement.

\begin{figure}
\plotone{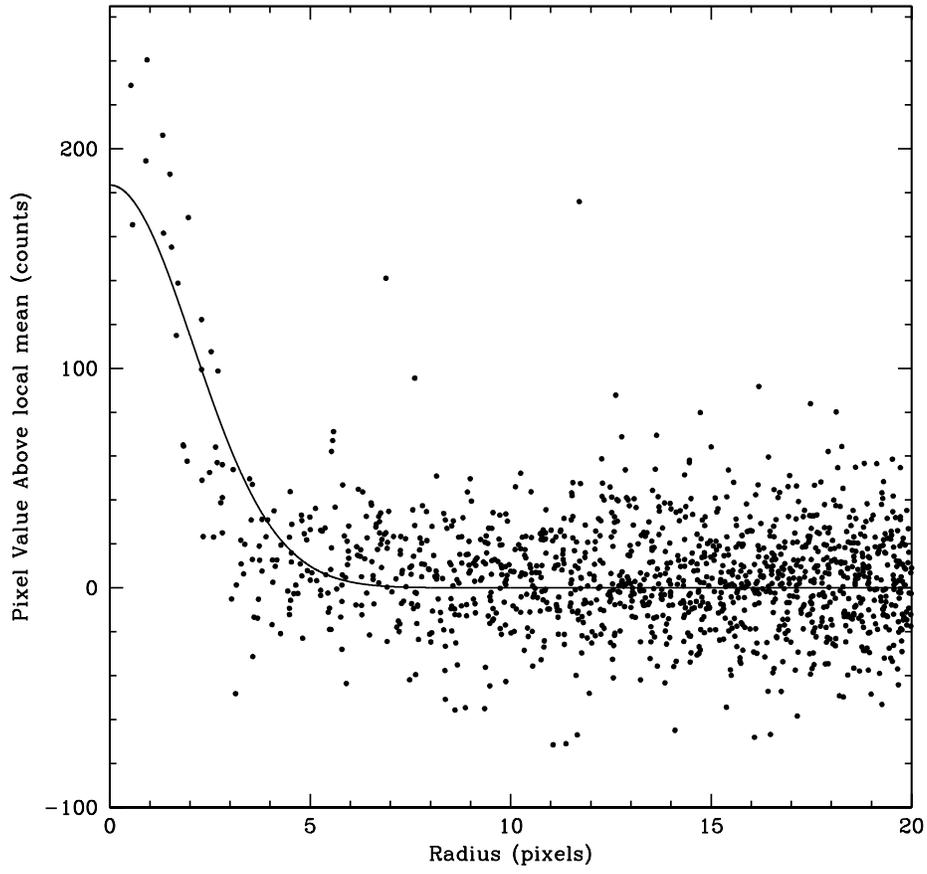}
\caption{Radial profile showing, for reference, the best Gaussian fit (amplitude $183.5$ and FWHM $4.87$). The data is steeper than a Gaussian.}
\label{fig:radial_profile}
\end{figure}

\subsection{Angular Size}
The light echo candidate is resolved in the STIS image and has a full-width half-maximum (FWHM) of $0.24\pm0.03$ arcsec. The radial profile is shown in Fig.~\ref{fig:radial_profile}. This diameter is small enough that the point-spread function (PSF) of the telescope contributes slightly to the image width. We can estimate the deconvolved echo radius by assuming it is a sharp-edged disk convolved with the STIS PSF which has a FWHM of about $0.06$ arcsec. This value is calculated from the intensity profile in the STIS manual. The angular size of the disk was varied until the convolved image was a good match to the observed echo width. Doing this we find that the deconvolved angular radius of the echo is \radiusvalue{} including error for the convolution process. If we assume the echo is an annulus we do not get a good match to the observed light distribution unless the radius of the inner hole is smaller than $0.03$ arcsec, which is roughly half a pixel. Although one of the innermost pixels is suggestively under-luminous which could be a hint of annulus structure, the resolution of the data was too low to confirm if it exists. A convolved disk of radius \radiusvalue{} was a good match to the STIS image.

\subsection{Dust Distance}
The geometry of the dusty region causing \SN{1995E}'s echo is an important unknown. A pioneering study by \citet{Couderc1939} investigated light echoes for several key geometries such as the flat dust sheet, which is one of the simplest and most important cases, and the dust cloud. An analytic treatment shows that the dust sheet produces an echo that is a circular ring containing the source and the radius of the ring at a given time does not depend on the inclination of the sheet but rather the distance of the supernova to the dust sheet. If the sheet is perpendicular with respect to the line-of-sight, the source lies at the center of the ring; otherwise the source is shifted off center. Such a shift would be small compared to the astrometric accuracy in this study. Furthermore, because the host galaxy is seen nearly face-on (cf.~Fig.~\ref{fig:closeup}) and the dust may be associated with the disk, it is plausible that the assumption of a perpendicular plane is not as restrictive as it may at first appear and is taken hereafter. The width of the ring depends on the thickness of the dust sheet, $\Delta{}R$, and the duration of the explosive event. The thickness and duration are assumed to be small and short. Under these conditions, the distance, $R$, of the supernova to the dust plane is given in approximation by
\begin{equation}
D^{2}\cong\frac{2c\,\Delta{}T\,R+(c\Delta{}T)^{2}}{\theta^{2}},
\end{equation}
where $c$ is the speed of light, $D$ is the distance to the host galaxy, $\theta$ is
the angular radius of the echo, and $\Delta{}T$ is the time since maximum. This approximate formula is good to better than $0.01\%$ for typical values compared to the exact expression. Hubble's Law, $D=v/H_{0}$, gives the distance to the host galaxy, where $H_0=70\pm4$ km s$^{-1}$ Mpc$^{-1}$ \citep{Tegmark2004}. The recessional velocity of \NGC{2441} is $v=3470\pm17$ km s$^{-1}$ \citep{Falco1999}. This yields a distance of \Rvalue{} to the dust sheet of \SN{1995E}. An annulus (or a solid disk, as observed in this case) instead of a thin ring would result from the breakdown of some of simplifying assumptions of this model such as a dust sheet that is non-negligibly thin or a finite duration of the source event.

\subsection{Brightness}
The flux, $f_{\lambda}$, and 50CCD STIS magnitude, CL, of the image are given by the STIS
Data Handbook as
\begin{equation}
f_{\lambda}=\frac{counts \times PHOTFLAM}{EXPTIME}
\end{equation}
and
\begin{equation}
CL=-2.5\log{f_{\lambda}}+PHOTZPT.
\end{equation}
The values of PHOTFLAM and PHOTZPT are given in the image headers as $9.9509750\times{}10^{-20}$ ergs s$^{-1}$ cm$^{-2}$ \AA{}$^{-1}$ and $-21.1$ mag respectively (which corresponds to a one count per second magnitude of
$26.41$). PHOTZPT is set such that the star Vega has a Johnson $V$-band magnitude of zero. The STIS image yields a count rate for the echo candidate of $4.37$ DN sec$^{-1}$
($3935$ counts) within a $6$-pixel radius aperture for a non-aperture corrected magnitude of $24.80\pm0.08$. The error is that reported by IRAF task ``phot''.

The target image is compact and an aperture correction should not be much different from that of a stellar source. An aperture correction can be found using the 50CCD energy-encircled radial profile (Table~14.3) of the STIS Instrument Handbook. Linear interpolation between $5$ and $10$ pixels is used to find the intensity at $6$-pixels. We chose $20$-pixels, where the encircled energy is $99.6$\%, as the outer, ``infinite aperture'' radius. The aperture correction is $\Delta{}m_{6\rightarrow20}\cong-0.093$ mag and so the aperture-corrected CL magnitude is $24.71\pm0.13$, which includes $0.05$ mag error estimated and added just for the aperture correction.

A transformation from the CL magnitude of the 50CCD detector with its large
bandpass to a Johnson $V$-band magnitude is given by combining Eqs.~1 and 2 of \citet{Rejkuba2000} and eliminating their 50CCD$-$F28X50LP color
variable. This yields the following equation:
\begin{equation}
V=CL+(0.26\pm0.19)\frac{(\colorindex{V}{I})-1.18\pm0.03}{1.76\pm0.13}+0.01\pm0.04.
\label{eq:rejkuba}
\end{equation}
The $V$-band magnitude thus depends on the \colorindex{V}{I} color.
We have no direct estimate of the color of the \SN{1995E} echo so it must be approximated. A formula connecting the unextincted color of a supernovae at maximum, $(\colorindex{B}{V})^{0}_{max}$, to the color of its echo is given by Eq.~20 of \citet{Patat2005} for a single-scattering plus attenuation (SSA) model. The same formula written in terms of the \colorindex{V}{I} color and rewritten in terms of the reddening instead of optical depth is
\begin{equation}
(\colorindex{V}{I})\approx(\colorindex{V}{I})^{0}_{max}-2.5 \log{\frac{C^{V}_{ext}\omega{}^{V}\Delta{}t^{V}_{SN}}{C^{I}_{ext}\omega{}^{I}\Delta{}t^{I}_{SN}}}+(R_V-R_I)E(\colorindex{B}{V}).
\label{eq:patatlike}
\end{equation}
The values of the dust cross-section, $C_{ext}$, albedo, $\omega$, and flash duration, $\Delta{}t_{SN}$, are shown in Table~\ref{tab:dust} for the $B$, $V$, and $I$-bands. The cross-section and albedo were calculated using the latest synthetic extinction models of Weingartner and Draine \citep{Weingartner2001,Draine2003}. For the flash duration, we used the $B$ and $V$ values reported in the \citet{Patat2005} paper, which were based on \SN{1992A} and \SN{1994D}. We measured the $I$-band value using the light curve of \SN{1994D} alone \citep{Patat1996}. The $I$-band light curves of Type~Ia supernovae can be quite heterogeneous because of the secondary maximum and this can affect the value of $\Delta{}t_{SN}^{I}$ so \textit{caveat emptor}. The flash durations are actually weak functions of $\Delta{}m_{15}(B)$ but they were treated as constant for our supernovae. Patat points out that the scattering term is small for the \colorindex{B}{V} analog of Eq.~\ref{eq:patatlike} and the echo color can be approximated by just the color of the supernovae at maximum plus a term linear in the optical depth. In \colorindex{V}{I}, the scattering term is non-negligible mostly because of the change in the ratio of the dust cross-section. An even more serious concern is the effect of multiple-scattering because \SN{1995E} has a large optical depth along the line of sight ($\tau{}_{d}\approx 2.1$). Comparison of Patat's SSA and Monte Carlo (MC) models show that multiple scattering can play a large role even in \colorindex{B}{V} for large $\tau{}_{d}$.

The $V$-band magnitude of the echo can now be found from Eqs.~\ref{eq:rejkuba} and \ref{eq:patatlike} given the extinction. But how accurate is this model? The only test case is \SN{1998bu}. The predicted value of its \colorindex{B}{V} color (using the \colorindex{B}{V} version of Eq.~\ref{eq:patatlike}) compared to our measured WFPC2 color (see below) shows about a $0.2$ mag difference. This is compatible with the RMS scatter found by \citet{Rejkuba2000} for the color-color equations used to derive Eq.~\ref{eq:rejkuba} and was assigned as the error in our \colorindex{V}{I} echo color. The $V$-band brightness of \SN{1995E}'s echo is $24.57\pm0.18$ mag using $R_{V}=3.1$. Using the disk previously discussed, the surface brightness is $21.21\pm0.25$ mag arcsec$^{-2}$. The STIS $V$-band magnitudes of \SN{1991T} and \SN{1998bu} were $22.34\pm0.23$ and $21.55\pm0.20$ mag, respectively.

Ratios of total-to-selective extinction of $R_{V}=3.1\pm0.1$ and $R_{I}$=$1.48$ were used above \citep{Cardelli1989}. This is to remain consistent with the lowest $R_{V}$ dust and echo models which also used $R_{V}=3.1$. However, it is believed that Type~Ia supernovae have $R_{V}\approx2.55\pm0.3$ \citep{Riess1996} or even as low as $1.80\pm0.19$ \citep{Elias-Rosa2006}. An $R_{V}$ of $2.55$ would make the STIS observations for \SN{1995E}, \SN{1991T}, and \SN{1998bu} fainter by $0.06$, $0.01$, and $0.03$ mag, respectively. The echo color for large optical depths is also affected by changes in $R_{V}$ but it is worth noting that the interstellar extinction law is less $R_{V}$-dependent in the $I$-band than the $B$-band. This mitigates some worries of the model dependencies on the dust properties but lower $R_{V}$ models would be very useful.

\begin{deluxetable}{ccccl}
\centering
\tablewidth{0pt}
\tablecaption{Dust parameters and flash durations}
\tablehead{\colhead{Filter} & \colhead{$\lambda$(\AA)} & \colhead{$C_{ext}(10^{-22}$cm$^{2})$} & \colhead{$\omega$} & \colhead{$\Delta{}t_{SN}$(yr)} } 
\startdata
B & 4300 & 6.59 & 0.65 & 0.0654 \\
V & 5400 & 4.98 & 0.67 & 0.0857 \\
I & 9000 & 2.36 & 0.64 & 0.102 \\
\enddata
\tablecomments{The values of extinction cross-section and albedo, $C_{ext}$ and $\omega$, for $R_{V}=3.1$ from \citet{Weingartner2001} and \citet{Draine2003}. The values of the flash durations, $\Delta{}t_{SN}$, are from \citet{Patat2005} except for $\Delta{}t^I_{SN}$ which was measured using the light curve of \SN{1994D}.}
\label{tab:dust}
\end{deluxetable}

The WFPC2 PC data set of \SN{1998bu} had the CR-splits combined and the resulting images for each visit and filter were averaged together (to remove the effects of sky quantization) and an instrumental magnitude was measured. The conversion of the F439W and F555W magnitudes to standard $B$ and $V$ magnitudes for the echo was accomplished using the method of \citet{Holtzman1995}. Their work describes the transformation of the WFPC2 flight system magnitudes with a $0\farcs{}5$ aperture to Johnson magnitudes. Our analysis is complicated by the extended nature of the source ($\approx{}0\farcs{}83$ radius ring); it demands a larger aperture but this will overestimate the flux. Our working solution was to use a large aperture and correct back the required $0\farcs{}5$ using the energy-encircled profiles in the Holtzman paper. It was necessary to use the color measured for the later visit in the calculation of the magnitude of earlier one since it was only imaged in F555W\@. We find $V=21.11\pm0.04$ on 2000-Jun-19 and $B=21.31\pm0.06$ and $V=21.24\pm0.05$ on 2002-Jan-10.

\section{Discussion}

Currently there are only two known Type Ia SNe light echoes from recent events: \SN{1991T} and \SN{1998bu} (see Table~\ref{tab:compare}). All three echoes have a late-type host galaxy (that is, $>$Sa). \SN{1991T} in \NGC{4527} was a very unusual SN~Ia. It showed weak spectral features and exhibited broad $B$ and $V$ light curves ($\Delta{}m_{15}(B)=0.94\pm0.05$). Later it showed a more normal spectrum and decline rate; however, at about $600$ days past maximum, the exponential fading slowed until the decline rate was consistent with zero and the spectrum showed features reminiscent of a supernova at maximum, an observation which was identified as the result of a light echo \citep{Schmidt1994} caused by a dust cloud of radius $50$ pc \citep{Sparks1999}. On the other hand, \SN{1998bu} in \NGC{3368} was spectroscopically and photometrically normal except for appearing to be more reddened than usual (see \citealt{Suntzeff1999}) until, at around $400$ days past maximum, the light curve started to deviate from the expected decline. By $700$ days past maximum it had almost flattened. \citet{Cappellaro2001} attribute this behavior to a light echo caused by foreground dust roughly $100$ pc away by comparing an integrated early spectrum to the late time spectrum. \citet{Garnavich2001} discovered from HST WFPC2 imaging that there is not one but two echoes for this event corresponding to dust at $120\pm15$ pc and less than $10$~pc away.

\begin{figure}
\plotone{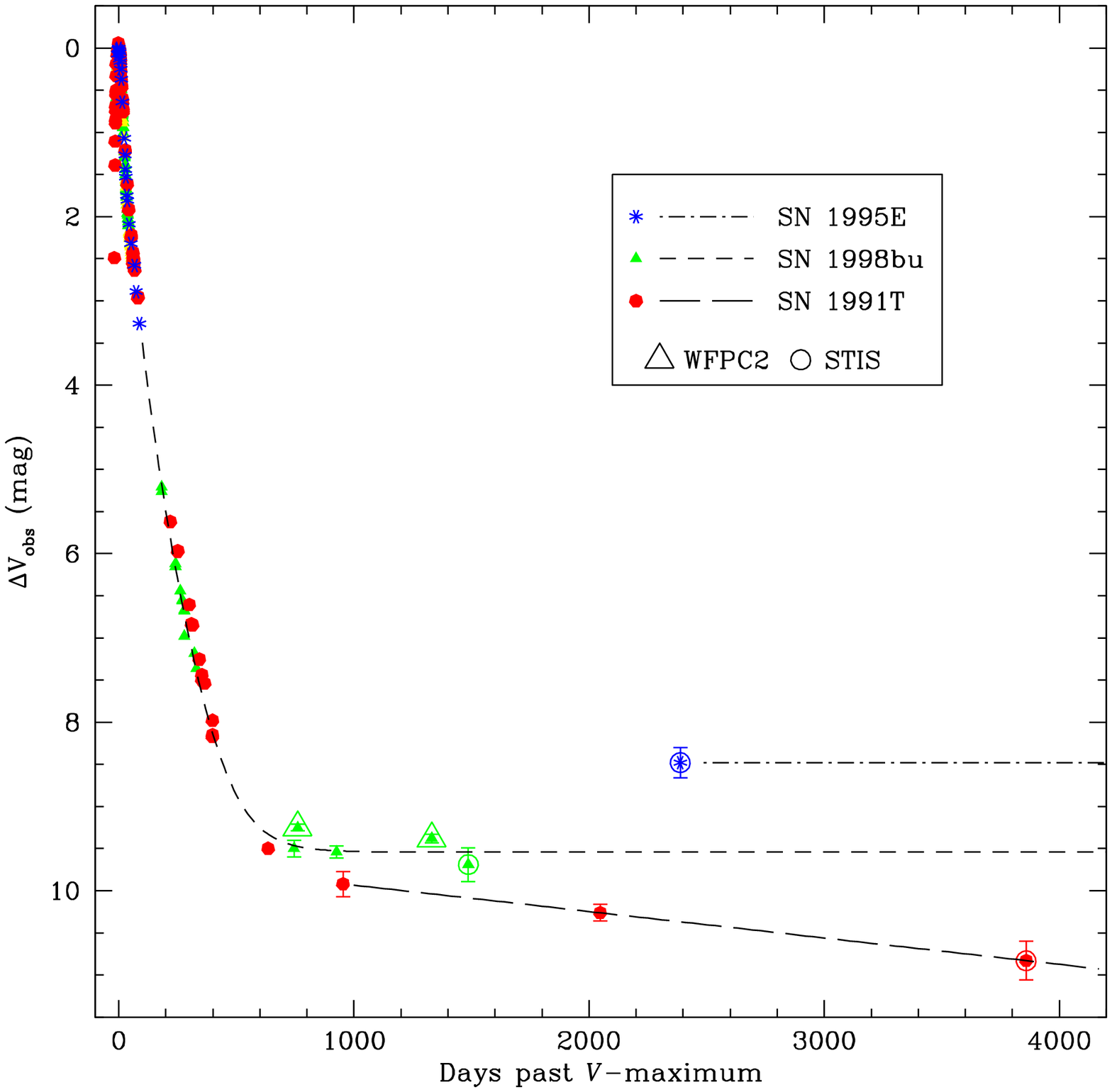}
\caption{Light curves of \SN{1995E}, \SN{1998bu}, and \SN{1991T} showing the decline in magnitude of the supernovae from $V_{max}$, $\Delta{}$V$_{obs}$, as a function of time after $V$-maximum. Clearly the echo of \SN{1991T} is fading ($\approx0.114$ mags per year) and perhaps also for \SN{1998bu} while \SN{1995E} is inconclusive. The ground-based light curve data are from \citet{Lira1998} and \citet{Schmidt1994} for \SN{1991T} (early and late-time points, respectively), from \citet{Riess1999} for \SN{1995E}, and from \citet{Jha1999} and \citet{Suntzeff1999} for \SN{1998bu}.}
\label{fig:lightcurves}
\end{figure}

How does the integrated brightness of \SN{1995E}'s echo compare to other Type Ia light echoes? The $V$-band magnitude difference between maximum and $\Delta{}T=2387$ days ($\sim{}6.5$ yrs) past maximum for \SN{1995E} was $8.48\pm0.18$ mag. The brightness differences for \SN{1991T} and \SN{1998bu} for the same time past maximum, as can be estimated from Fig.~\ref{fig:lightcurves}, are $10.37\pm0.23$ and $9.54\pm0.20$ mag, respectively. The data clearly indicate the light curve of \SN{1991T} is declining nearly linearly in magnitude at a rate of about $0.114$ mags per year. The light curve of \SN{1998bu} is more difficult to interpret because of the two echoes and the baseline for the observations after the break from the normal exponential-like decline is shorter. Linear regression on its last points both with and without weighting indicate a slope similar to that of \SN{1991T} but the error in the slope is relatively large and moreover these points are influenced by the elbow in the light curve. These things considered, the slope is consistent with zero. The models of Patat show that the luminosity of a dust sheet echo declines at a much slower rate on average than a dust cloud (roughly $0.03$--$0.04$ versus $0.11$--$0.14$ magnitudes per year depending on the optical depth); so the cloud model matches the decline rate of \SN{1991T} quite well. The dust sheet model for the outer echo of \SN{1998bu} seems strongly motivated but the evidence suggests the possibility of a dust cloud for the inner echo. Our data does not seem inconsistent with the inner echo being produced by a dust cloud but is too ambiguous to draw a firm conclusion from because of the short baseline of observation. It was therefore assumed to plateau as might be expected of an echo produced by a dust sheet.

\begin{deluxetable}{lllccc}
\centering
\tablewidth{0pt}
\tablecaption{Type Ia light echoes}
\tablehead{\colhead{Event} & \colhead{JD($V_{max}$)} & \colhead{$V_{max}$(mag)} & \colhead{$\Delta{}m_{15}(B)$} & \colhead{Host}} 
\startdata
\SN{1991T}  & $2448378.3^a$ & $11.51(02)^a$  & $0.94(05)^c$ & \NGC{4527}\\
\SN{1995E}  & $2449776.4$   & $16.09^d(02)$  & $1.06(05)^c$ & \NGC{2441}\\
\SN{1998bu} & $2450954.4^b$ & $11.86(02)^b$  & $1.01(05)^c$ & \NGC{3368}\\
\enddata
\tablerefs{(a) \citet{Lira1998} (b) \citet{Jha1999} (c) \citet{Phillips1999} (d) \citet{Riess1999}.}
\tablecomments{The date of maximum for \SN{1995E} was calculated using a parabolic fit to the brightest three data points of \citet{Riess1999}. Its $V_{max}$ error was assigned for this paper.}
\label{tab:compare}
\end{deluxetable}

It is useful to account for the effects of extinction such that the relative brightness of each echo to the supernova at maximum can be compared. Unfortunately, the line-of-sight extinction for arbitrary dust geometries does not correlate with the effective extinction of the echo (e.g., if a dust sheet has a hole directly in front of a SN there would be less line-of-sight extinction); but to make progress, the line-of-sight extinction was taken to be approximately equal to the effective extinction of the echo. Table~\ref{tab:decline} details how this was accomplished. The values of host galaxy extinction, $E(\colorindex{B}{V})$, given in the table are those of \citet{Phillips1999} (called $E(\colorindex{B}{V})_{Avg}$ there). They have removed the effects of Galactic reddening using \citet{Schlegel1998} and included the K-correction to first order. Since we want the difference in the brightness of the supernovae to the echo, the Galactic extinction cancels. After correcting for extinction (an $R_{V}$-value of $3.1$ and the single scattering approximation, where there is no extinction along the echo light path, were used), all the observed echoes are about $10.5$ to $11.0$ mag fainter than the supernova at maximum. This narrow range is somewhat striking given the diversity of the environments that could have produced the echoes. For the dust sheet approximation the echo brightness is a fairly sensitive function of the dust distance and closer dust results in a brighter echo. The three echoes here have a range of dust distances from roughly $200$ to $700$~lyr but show a narrow range in brightness.

\begin{figure}
\plotone{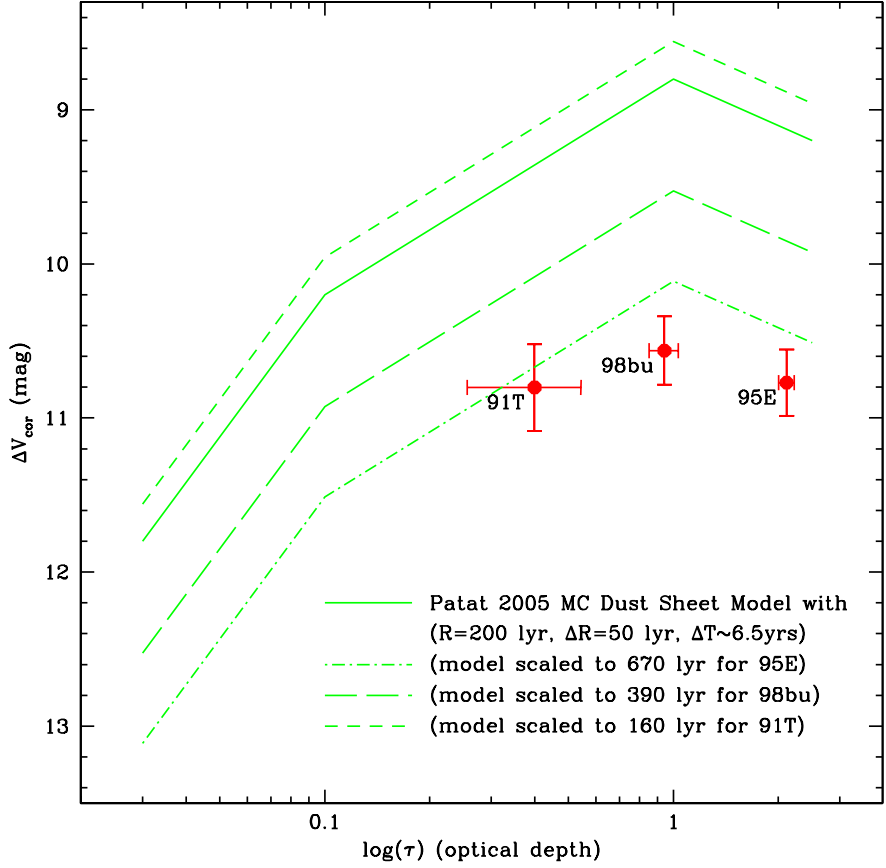}
\caption{Decline in magnitude (corrected for extinction) at $\Delta{}T=2387$ days vs the optical depth, $\tau{}=A_{V}/(2.5 \log{e})$. The solid-line data points come from the Fig.~$6$ Monte-Carlo solutions of \citet{Patat2005} and the dashed-line points are the same model scaled via Patat's single-scattering formula $4\pi{}d^{2}F(t)/L_{0}\approx0.3\tau{}_{d}/R$ to the dust distances of each supernova (rounded to the nearest $10$~lyrs). The effect of a dust mixture change to $R_{V}=2.55$ is discussed in the text.}\label{fig:tau}
\end{figure}

The extinction-corrected echo luminosities of \SN{1995E}, \SN{1998bu}, and \SN{1991T} are plotted against scattering optical depth in Fig.~\ref{fig:tau}. Also plotted are the Monte-Carlo simulations of Patat for a dust sheet of thickness $50$~lyr ($\sim15.3$~pc) at $200$~lyr ($\sim61.3$~pc) (the solid line) and the same solution scaled to the dust distance of each supernova. The luminosity of the echoes were assumed to scale inversely proportional to the dust distance via the single-scattering approximation formula, $4\pi{}d^{2}F(t)/L_{0}\approx0.3\tau{}_{d}/R$ \citep{Patat2005}, for a perpendicular sheet. Here $d$ is the distance from the observer to the SN, $F$ is the observed integrated flux of the echo, $L_{o}$ is the luminosity of the SN treating it as a flash of duration $\Delta{}t_{SN}$, $\tau{}_{d}$ is the optical depth, and $R$ is the distance to the plane. The distance to \SN{1995E}'s dust sheet as determined by this paper (the dashed-dotted line) is \Rvalue{} ($\sim670$~lyr). The predicted luminosity of \SN{1995E}'s echo for its optical depth agrees with our data within the statistical and a reasonable model uncertainty. \SN{1991T} had a light echo consistent with uniform density dust cloud of radius $50$~pc ($\sim160$~lyr) \citep{Sparks1999} and is over a magnitude fainter than one would expect ($\approx1.1$~mag). Again, the case of \SN{1998bu} is more complicated because there are two echoes. The model curve shown in Fig.~\ref{fig:tau} has been shifted to the $120$~pc ($\sim390$~lyr) distance corresponding to the outer echo. The echo luminosity implied is also around a magnitude fainter than the model predicts ($\approx0.9$~mag). Both these points appear to be under-luminous for a thin-sheet model but in both cases dust clouds, not dust sheets, are known or suspected to be involved and very large optical depths suggesting a full MC treatment may be needed for these echoes. Changing the dust mixture to $R_{V}=2.55$ has a significant effect on the values of $A_{V}$ that amounts to $-0.12$, $-0.41$, and $-0.18$ mag for \SN{1991T}, \SN{1995E}, and \SN{1998bu}, respectively. The $0.06$ mag discussed earlier for the STIS $V$-band magnitude of \SN{1995E} also directly affects $\Delta{}V_{cor}$ because there is only one late-time observation. The model now closely matches \SN{1995E} but the other two are still under-luminous. It will be interesting to see how the next echo discoveries compare to these.

\begin{deluxetable}{lrcrc}
\centering
\tablewidth{0pt}
\tablecaption{Extinction-corrected declines}
\tablehead{\colhead{Event} & \colhead{$\Delta{}V_{obs}$(mag)} & \colhead{$E(\colorindex{B}{V})$(mag)} & \colhead{$A_{V}$(mag)} & \colhead{$\Delta{}V_{cor}$(mag)}}
\startdata
\SN{1991T}  & 10.37(0.23) & 0.14(0.05) & 0.434(0.16) & 10.80(0.28) \\
\SN{1995E}  &  8.48(0.18) & 0.74(0.03) & 2.294(0.12) & 10.77(0.22) \\
\SN{1998bu} &  9.54(0.20) & 0.33(0.03) & 1.023(0.10) & 10.56(0.22) \\
\enddata
\tablecomments{This table calculates the decline of the echo relative to the supernova at maximum corrected for extinction, $\Delta{}V_{cor}=\Delta{}V_{obs}+A_{V}$, where as usual $A_V=R_V E(\colorindex{B}{V})$. The ratio of total to selective extinction was $R_{V}=3.1\pm0.1$ and the extinction values of the host galaxy, $E(\colorindex{B}{V})$, are those of \citet{Phillips1999}. For \SN{1995E} and \SN{1998bu}, the error in $\Delta{}V_{obs}$ is quite model dependent and these are estimates.}
\label{tab:decline}
\end{deluxetable}

\clearpage

\section{Conclusion}

We show that a resolved source resides at the position of Type Ia \SN{1995E} in face-on spiral \NGC{2441} which is likely to be a light echo. The angular size and brightness are consistent with an echo interpretation for the source; however, further high-resolution imaging is required to test for variability in size or brightness. If it is an echo, we derive a distance of \Rvalue{} between the scattering dust and the supernova.

Comparison of the \SN{1995E} echo with echoes observed from \SN{1991T} and \SN{1998bu} shows a strong similarity in their integrated flux relative to the peak supernova brightness despite large variations in the dust optical depth and dust distributions. Photometry of the \SN{1991T} echo suggests a decline in integrated brightness which is consistent with a cloud or shell distribution of dust around the progenitor. A new echo strengthens the case that some Type Ia SNe are near dusty environments and that a subset may be linked to a younger population. Further studies of SN~Ia light echoes are needed and may provide important constraints on progenitor models and any connection between progenitors and dusty or star-forming regions.

\acknowledgements{This research has made use of the NASA/IPAC Extragalactic Database (NED) which is operated by the Jet Propulsion Laboratory, California Institute of Technology, under contract with the National Aeronautics and Space Administration.}

\end{document}